\documentclass[pra, 12 pt]{revtex4}
\usepackage[english]{babel}
\usepackage{amsmath}
\usepackage{epsfig}

\begin{document}

\title{\bf On anomalous diffusion in a plasma in velocity space}
\author{S. A. Trigger $^{1}$, W. Ebeling $^2$, G.J.F. van Heijst $^3$, P.P.J.M. Schram $^3$, I.M. Sokolov $^2$}
\address {$^1$ Joint Institute for High Temperatures, Russian Academy of
Sciences, Izhorskaya 13/19, 125412 Moscow, Russia; e-mail: satron@mail.ru\\
$^2$ Institut f\"ur Physik, Humboldt-Universit\"at zu Berlin,
 Newtonstra{\ss}e 15, D-12489 Berlin, Germany;\\
 $^3$ Eindhoven  University of Technology, P.O. Box 513, MB 5600
Eindhoven, The Netherlands}
\date{12 January 2010}

\begin{abstract}

The problem of anomalous diffusion in momentum space is
considered for plasma-like systems on the basis of a new
collision integral, which is appropriate for consideration of the
probability transition function (PTF) with long tails in 
momentum space. The generalized Fokker-Planck equation for description
of diffusion (in momentum space) of particles (ions,
grains etc.) in a stochastic system of light particles
(electrons, or electrons and ions, respectively) is applied to the
evolution of the momentum particle distribution in a plasma. In a plasma the developed approach is
also applicable to the diffusion of particles with an arbitrary mass relation, due to the small
characteristic momentum transfer. The
cases of an exponentially decreasing in momentum space (including the Boltzmann-like) kernel in the
PT-function, as well as the more general kernels, which create the
anomalous diffusion in velocity space due to the long tail in the
PT-function, are considered.
Effective friction and diffusion coefficients for plasma-like systems are found.\\

PACS number(s): 52.27.Lw, 52.20.Hv, 05.40.-a, 05.40.Fb

\end{abstract}

\maketitle

\section{Introduction}

Diffusion in coordinate and in momentum (velocity) space is of
fundamental importance and has attracted a growing interest during
many years, since the description of this processes provides a
simplified and effective key for addressing of many problems of
kinetic theory.

Deviations from a linear time-dependence, i.e. $<r^2(t)>\sim t$, 
of the mean square displacement in \emph{coordinate space} has
been experimentally observed, in particular, under essentially
non-equilibrium conditions or for some disordered systems. The
average square separation of a pair of particles passively moving
in a turbulent flow grows, according to Richardson's law, with the
third power of time [1]. For diffusion typical for glasses and
related complex systems [2], the observed time dependence is slower
than linear. These two types of anomalous diffusion are obviously 
characterized as superdiffusion $<r^2(t)>\sim t^\alpha$
$(\alpha>1)$ and subdiffusion $(\alpha<1)$ [3]. For a description
of these two diffusion regimes a number of effective models and
methods have been suggested. The continuous time random walk
(CTRW) model of Scher and Montroll [4], leading to
subdiffusion behavior, provides a basis for understanding
photoconductivity in strongly disordered and glassy
semiconductors. The Levy-flight model [5], leading to
superdiffusion, describes various phenomena as self-diffusion in
micelle systems [6], reaction and transport in polymer systems [7],
and is applicable even to the stochastic description of financial
market indices [8]. Both cases can be effectively described
by generalized diffusion equations with fractional derivatives in time or in
coordinates, respectively [9]. For example, different
aspects of the anomalous diffusion in coordinate space were
considered within this scheme in Refs. [10], [11].

However, recently a more general approach has been
suggested in [12], [13]. This one permits to reproduce the results of
the standard fractional differentiation method in coordinate
space, when the latter is applicable, and enable to describe
more complicated cases of anomalous diffusion processes. In [14]
this approach has been applied also to the case of diffusion in a
time-dependent external field in coordinate space. In what
follows, we concentrate on the problem of diffusion in momentum space
in application to plasma systems.

Problems of diffusion in \emph{momentum space} have been
considered for plasmas in the fundamental study by Landau [15] and
later by Rostoker and Rosenbluth, Lenard, Balescu, Klimontovich
and many others. Various theoretical and experimental aspects
of these investigations can be found in [16-18].

Here our main interest is focused on anomalous diffusion in
momentum space by using the methods developed in [12],[13] for
coordinate space. Recently, see [19], a new kinetic equation for
anomalous diffusion in velocity space has been derived on the
basis of an appropriate expansion of PTF and some particular
problems were investigated on this basis. In the present paper the
problem of anomalous diffusion in momentum (velocity) space
will be considered for plasma-like systems.

Some aspects of anomalous diffusion in velocity space have
been investigated during the last decade in a number of studies [20-23]. In
particular in [22] the phenomenological equation for anomalous diffusion in
velocity space for a magnetized plasma has been obtained on the
basis of the Langevin model with the linear friction proportional
to the particle velocity and with non-Gauss noise. The
corresponding Fokker-Planck type equation included a diffusion term
with a fractional derivative and the usual drift term with
the first momentum derivative. The applicability of this equation
is limited, due to phenomenological nature of the considered model. We should 
mention the analogy of this equation with the
structure of the equation for anomalous diffusion in coordinate
space under action of an external field [11],[14]. An equation similar to
that in [22] has recently been  applied in [23] to describe the evolution
of the velocity distribution function of strongly non-equilibrium
hot and rarified plasmas. Such types of plasma exist, e.g., in tokomaks.

On the whole, in comparison with anomalous diffusion in
coordinate space, anomalous diffusion in velocity space is
has been studied to a modest extent.

In this paper the problem of anomalous diffusion in momentum
(velocity) space will be considered for plasma-like systems.
In spite of formal similarity, the physical (and mathematical)
nature of diffusion in momentum space is very different from
that in coordinate space. This is clear already from the fact
that momentum conservation, which takes place in momentum
space, has no analogy in coordinate space.

Diffusion in velocity space for the cases of normal and anomalous
behavior of the PT-function is presented in Section II.
Starting from the argumentation based on the Boltzmann type of the
PTF, we describe the new approach to the kinetic equation, which
in fact can be applied to the wide class of PTF functions
based on the prescribed distribution function for one (light) sort
of particles. The anomalous diffusion in momentum space for plasma
is analyzed in Section III on the basis of the Boltzmann-type
kernel for PTF. Models of anomalous diffusion for plasma-like
systems are considered in Section IV. In Section V the generalized
Fokker-Planck equation for diffusion, written for the
formal Fourier-component $f ({\bf s},t)$ of the distribution
function $f ({\bf p},t)$, is represented in partial derivatives
in velocity space. This representation is possible only in particular
cases of the power dependence of the coefficients in
the generalized diffusion equation.

\section{Diffusion in the velocity space on the basis of
a master-type equation}

Let us consider now the main problem formulated in the
introduction, namely, diffusion in momentum space ($V$-space) on
the basis of the master equation, which describes the
balance of grains entering and leaving point ${\bf p}$ at time $t$ (see, e.g. [24,25])
\begin{equation}
\frac{df({\bf p},t)}{dt} = \int d{\bf q} \left\{w ({\bf p, p'})
f({\bf p'}, t) - w ({\bf p', p}) f({\bf p},t) \right\}.
\label{DC2aa}
\end{equation}
The structure of this equation is formally similar to the master
equation (see, e.g. [13]) in coordinate space.
Here $w ({\bf p, p'})$ is the kernel describing the transition
probabilities. Note that there is only one rather general
condition which $w ({\bf p, p'})$ should satisfy if the
stationary solution exists: the balance condition of the detailed 
for a stationary distribution function $f^{st}({\bf p})$, which reads
\begin{equation} \frac{w({\bf p,p'})}{w({\bf p',p})}=
\frac{f^{st}({\bf p})}{f^{st}({\bf p'})}\label{DC2ab}
\end{equation}

In the following analysis we use a form of the master equation [26] equivalent to (\ref{DC2aa})
\begin{equation}
\frac{df({\bf p},t)}{dt} = \int d{\bf q} \left\{W ({\bf q, p+q})
f({\bf p+q}, t) - W ({\bf q, p}) f({\bf p},t) \right\}.
\label{DC2b}
\end{equation}
The probability transition $W({\bf p, p'})$ describes the probability
for a grain with momentum ${\bf p'}$ at point ${\bf p'}$ in momentum space to
transfer from this point ${\bf p'}$ to the point ${\bf p}$ per
unit time. The momentum transfer is equal ${\bf q= p'- p}$. Of course, as mentioned above, the overall momentum has to be conserved.

Assuming in the beginning that the characteristic changes in momentum
are small one may expand Eq.~(\ref{DC2b})  and arrive at the
Fokker-Planck form of the equation for the density distribution
$f({\bf p},t)$
\begin{equation}
\frac{df({\bf p},t)}{dt} = \frac {\partial}{\partial p_\alpha}
\left[ A_\alpha ({\bf p}) f({\bf p},t) + \frac{\partial}
{\partial p_\beta} \left(B_{\alpha\beta}({\bf p}) f({\bf p},t)
\right)\right], \label{DC3b}
\end{equation}
\begin{equation}
A_\alpha({\bf p}) = \int d^r q q_\alpha W({\bf q, p});\;\;\;\
B_{\alpha\beta}({\bf p})= \frac{1}{2}\int d^r q q_\alpha q_\beta
W({\bf q, p}). \label{DC4b}
\end{equation}
The coefficients $A_\alpha$ and $B_{\alpha \beta}$ describe the
friction force and diffusion, respectively. Here $r$ is the
momentum space dimension.

Because the velocity of heavy particles is small, the $\bf
p$-dependence of the PTF can be neglected for the calculation of 
diffusion coefficient, which in this case is constant
$B_{\alpha\beta}=\delta_{\alpha\beta}B$, where B is the integral
\begin{equation}
B = \frac{1}{2r}\int d^r q q^2 W(q). \label{DC6b}
\end{equation}

If we neglect the $\bf p$-dependence of the PTF at all, we arrive
to the coefficient $A_\alpha=0$ (while the diffusion coefficient
is constant). In this approach, which is known to be incorrect, the
coefficient $A_\alpha$ for the Fokker-Planck equation can be
determined on the basis of the argument that the stationary distribution
function is Maxwellian. In this way we arrive at the standard form
of the coefficient $MT A_\alpha(p)=p_\alpha B$, which is one of
the forms of Einstein's relation. For systems far from
equilibrium this argument is not acceptable.

Following [1], we now generalize, the Fokker-Planck
approach to find the coefficients of the kinetic equation,  which
are applicable also to slowly decreasing PT functions. We apply a
more general approach, based on the difference of the velocities for
light and heavy particles. For calculation of the function
$A_\alpha$ we have take into account that the function $W(\bf
{q,p})$ is scalar and depends on the variables $q, {\bf q \cdot
p}, p$. Expanding $W(\bf {q,p})$ on $\bf {q\cdot p}$ one arrives at the following
approximate representation of the function $W(\bf {q,p})$:
\begin{eqnarray}
W({\bf q,p)}\simeq W(q)+\tilde W'(q)({\bf q \cdot p})+
 \frac{1}{2}\tilde W''(q) ({\bf q \cdot p})^2, \label{DC7b}
\end{eqnarray}
where $\tilde W'(q)\equiv \partial W (q, {\bf q \cdot p})
/\partial ({\bf q p})\mid_{{\bf q \cdot p}=0}$ and $\tilde
W''(q)\equiv \partial^2 W (q, {\bf q \cdot p}) /\partial ({\bf q
p})^2 \mid_{{\bf q \cdot p}=0}$.

Then, with the necessary accuracy, $A_\alpha$ equals
\begin{equation}
A_\alpha({\bf p}) = \int d^r q q_\alpha q_\beta p_\beta \tilde
W'(q)= p_\alpha \int d^r q q_\alpha q_\alpha  \tilde
W'(q)=\frac{p_\alpha}{r} \int d^r q q^2  \tilde W'(q)\label{DC10b}
\end{equation}
If the function $W({\bf q,p)}$ satisfies the equality $\tilde W'(q)=
W(q)/ 2 MT$, we arrive at the usual Einstein
relation
\begin{equation}
M T A_\alpha({\bf p}) =  p_\alpha B \label{DC11b}
\end{equation}

Let us check this relation for Boltzmann-type collisions,
which are described by the PT-function $W({\bf q, p)}=w_B({\bf q,
p})$ [13]:
\begin{eqnarray}
w_B({\bf q, p})=\frac{2\pi}{\mu^2 q} \int_{q/2\mu}^\infty du\,u\,
\frac{d \sigma}{do} \left[\arccos \, (1-\frac{q^2}{2\mu^2 u^2}), u
\right] f_b (u^2+ v^2-{\bf q \cdot v} /\mu), \label{DC12b}
\end{eqnarray}
where (${\bf p}=M{\bf v}$) and $d \sigma (\chi,u)/ do\, $, $\mu$ and $f_b$ are
the differential scattering cross-section, the mass and
distribution function for the light particles, respectively.
In Eq.~(\ref{DC12b}) we took into account
the approximate equalities for the scattering of light and heavy particles $q^2 \equiv
(\triangle {\bf p})^2=p'^2(1-cos\theta)$ and $\theta\simeq \chi$, where $p'=\mu u$ is the momentum of the light particle before collision.

For the equilibrium Maxwellian distribution $f_b^0$ the equality $\tilde W'(q)= W(q)/
2 MT$ is evident and we arrive at the usual Fokker-Planck equation
in velocity space with constant diffusion and friction coefficients $D \equiv B /M^2$
and friction $\beta \equiv B/MT=DM/T$, respectively, which satisfy
the Einstein relation.

For some non-equilibrium situations the PTF, as a function of the variable $q$, possesses a long
tail. In this case we have to derive a generalization of the kinetic
equation in the spirit of the analysis of the coordinate case
[13,14], because the diffusion and friction coefficients in the
form of Eqs.~(\ref{DC6b}) and (\ref{DC10b}) diverge for large $q$ if the
functions have an asymptotic behavior $W(q)\sim 1/q^\alpha$ with
$\alpha\leq r+2$ and (or) $\tilde W'(q)\sim 1/q^\beta$ with $\beta
\leq r+2$.

Inserting the expansion (\ref{DC7b}) for $W({\bf q, p})$ in
Eq.~(\ref{DC2b}) (as an example we choose $r=3$; the analysis for arbitrary $r$
runs in a similar way) we arrive at a new collision term of the kinetic
equation, which can be considered as a generalization of the Fokker-Planck equation
for anomalous diffusion in velocity space [19]:

\begin{eqnarray}
\frac{df({\bf s},t)}{dt} = A(s)f ({\bf s})+ B_\alpha
(s)\frac{\partial f ({\bf s},t)}{\partial {\bf s}_\alpha}
\label{DC16b}
\end{eqnarray}
In fact, as shown in [19], in the expansion (\ref{DC7b})
for $W({\bf q, p})$  we have to keep (with the necessary accuracy)
only the terms linear in ${\bf q p}$ and ${\bf p}$. The function
$f ({\bf s})$ in (\ref{DC16b}) is the Fourier-component $f
({\bf s})=\int \frac{d{\bf p}}{(2\pi)^3} \exp(i{\bf p s})f_g ({\bf
p},t)$ and the coefficients are equal to
\begin{eqnarray}
A(s)= \int d{\bf q} [\exp(-i{\bf(q s)})-1]W(q) = 4\pi
\int_0^\infty dq q^2 \left[\frac{sin\, (q s)}{qs}-1\right]W(q),
\label{DC17b}
\end{eqnarray}
\begin{eqnarray}
B_\alpha\equiv s_\alpha B(s);\;B(s)=-\frac{i}{s^2} \int d{\bf q}
{\bf q s} [\exp(-i{\bf(q s})-1]  \tilde W'(q)= \nonumber\\
\frac{4\pi}{s^2} \int_0^\infty dq q^2 \left[cos\, (q s)-\frac{sin
(q s)}{q s}\right]\tilde W'(q). \label{DC19b}
\end{eqnarray}
Here we took into account the existence of the small parameter
$~\mu/M$ and we omitted the small on this parameter terms of order $~{\bf p}^2$ and $W''$ in
Eq.~(\ref{DC7b}).

For the isotropic function $f({\bf s})=f(s)$ one can rewrite
Eq.~(\ref{DC16b}) in the form
\begin{eqnarray}
\frac{df(s,t)}{dt} = A (s) f_g (s)+ B(s)s \frac{\partial f
(s,t)}{\partial s}.  \label{DC22b}
\end{eqnarray}

For the case of strongly decreasing PTF the exponent under the
integrals for the functions $A(s)$ and $B(s)$ can be expanded as
\begin{eqnarray}
A(s)\simeq=-\frac{s^2}{6}\int d{\bf q}\, q^2 W(q);\; B(s)\simeq -
\frac{1}{3} \int d{\bf q} \, q^2 \tilde W'(q) \label{DC23b}
\end{eqnarray}

Then the simplified kinetic equation for the case of short-range
on $q$-variable PTF (non-equilibrium, in general case) reads
\begin{eqnarray}
\frac{df(s,t)}{dt} = A_0 s^2 f(s)+ B_0 s \frac{\partial f
(s)}{\partial s} ,\label{DC25b}
\end{eqnarray}
where $A_0\equiv -1/6 \int d{\bf q}\, q^2 W(q)$ and $B_0\equiv
-1/3 \int d{\bf q} \, q^2 \tilde W'(q)$.

The stationary solution of Eq.~(\ref{DC22b}) reads
\begin{eqnarray}
f(s,t)=C exp\,\left[-\int_0^s ds'\frac{A(s')}{s' B(s')}\right]
=C exp\,\left[-\frac{A_0 s^2}{2 B_0}\right] \label{DC26b}
\end{eqnarray}
The corresponding normalized stationary momentum distribution is given by
\begin{eqnarray}
f(p)=\frac{N_g B_0^{3/2}}{(2 \pi A_0)^{3/2}}\exp \,[-\frac{B_0
p^2}{2 A_0}]. \label{DC28b}
\end{eqnarray}
Therefore, in Eq.~(\ref{DC26b}) the constant $C=N$, where $N$ is the total number of particles in the system, which undergo diffusive motion. Equation
(\ref{DC25b}) and this distribution are the generalization of the
Fokker-Planck case for normal diffusion in a non-equilibrium
situation with strongly decreasing kernels $W(q)$, $W'(q)$ , when the prescribed
PTF function $W({\bf q,p})$ is determined,
e.g., by some non-Maxwellian distribution of the small particles
$f_b$. To show this in an alternative way let us take the Fourier
transformation of (\ref{DC16b}) and the corresponding coefficients $A$
and $B_\alpha$:
\begin{eqnarray}
\frac{df({\bf p},t)}{dt} = - A_0 \frac{\partial^2 f({\bf p},t)}{\partial p^2}
- B_0 \frac{\partial (p_\alpha f
({\bf p},t))}{\partial p_\alpha} ,\label{DC29b}
\end{eqnarray}
We then arrive at a Fokker-Planck type equation with
friction coefficient $\beta\equiv-B_0$ and diffusion coefficient
$D=-A_0/M^2$. In general these coefficients (Eq.~(\ref{DC23b})) do
not satisfy the Einstein relation.

In the case of equilibrium $W$-function (e.g., $f_b=f_b^0$, see
above) the equality $\tilde W'(q)=W(q)/2M T_b$ is satisfied. Then we find
$A(s)/s B(s)\equiv A_0/B_0 =M T_b$. In this case the
Einstein relation between the diffusion and friction coefficients $D=\beta T/M$
is satisfied and the standard Fokker-Planck equation is valid.

In the general case, however, the general equation (\ref{DC16b}), (\ref{DC17b}) and (\ref{DC19b})  have to be used.

\section{Diffusion in plasma systems on the basis of Boltzmann-type collisions}

Let us calculate the PTF for the case of Coulomb collisions. The
differential cross-section for the Coulomb scattering $d
\sigma/do$ equals
\begin{eqnarray}
\frac{d \sigma (\chi, u)}{do}=\left(Ze^2/2\mu u^2\right)^2 \frac{1}{sin^4 \frac{\chi}{2}}=\left(Ze^2/2\mu u^2\right)^2
\frac{16\mu^4u^4}{q^4}, \label{DC57b}
\end{eqnarray}
where $\chi=arccos(1-q^2/2 \mu^2u^2)$.
Then
\begin{eqnarray}
w_B^{Coul}({\bf q, p})=\frac{2\pi}{\mu^2 q} \int_{q/2\mu}^\infty du\,u\,\left(Ze^2/2\mu u^2\right)^2 \frac{16\mu^4u^4}{q^4}
f_b (u^2+ v^2-{\bf q \cdot v} /\mu)\nonumber\\
=\frac{8 Z^2 e^4\pi}{q^5} \int_{q/2\mu}^\infty du\,u\,
 f_b (u^2+ v^2-{\bf q \cdot v} /\mu). \label{DC58b}
\end{eqnarray}
It is necessary to stress that in the case of Coulomb
interaction the general equations (\ref{DC16b}),(\ref{DC22b}) are applicable not only for diffusion
of heavy particles in a light particle medium, but for arbitrary mass relations. The reason for this statement
is the typical small transfer of momenta in the Coulomb systems.

Let us calculate now the coefficients $A_\alpha ({\bf p}) $ and
$B_{\alpha\beta}({\bf p})$ to compare the results with the
linearized Landau kinetic equation, in which these coefficients
depend on ${\bf p}$. This implies that for the Coulomb interaction the
expansion by $({\bf q \cdot p})$ has to be performed at finite
${\bf p}$.

At first we consider the approximation in the spirit of the usual
Fokker-Planck approach. Eqs.~(\ref{DC4b}),(\ref{DC10b}) lead to the
expressions
\begin{equation}
A_\alpha({\bf p}) = \int d^3 q q_\alpha w_B^{Coul}({\bf q,
p})\cong \frac{p_\alpha}{3} \int d^3 q q^2  \tilde
w_B^{'\;\,Coul}(q),\label{DC59b}
\end{equation}
\begin{equation}
B_{\alpha\beta}({\bf p})= \frac{1}{2}\int d^3 q q_\alpha q_\beta
w_B^{Coul}({\bf q, p})\cong \frac{1}{2}\int d^3 q q_\alpha q_\beta
w_B^{Coul}(q), \label{DC60b}
\end{equation}
where
\begin{equation}
w_B^{Coul}(q)=\frac{8 Z^2 e^4\pi}{q^5} \int_{q/2\mu}^\infty
du\,u\, f_b (u^2),\label{DC61b}
\end{equation}
\begin{equation}
\tilde w_B^{'\;\,Coul}(q)=-\frac{8 Z^2 e^4\pi}{M\mu q^5}
\int_{q/2\mu}^\infty du\,u\, f_b' (u^2),\;\; \tilde
w_B^{'\;\,Coul}(q)=-\frac{4 Z^2 e^4\pi}{M T  q^5}
\int_{q/2\mu}^\infty du\,u\, f_b (u^2)\label{DC62b}
\end{equation}
and $f_b' (y) \equiv \partial f_b (y)/\partial y$. The procedure
which we used here implies that the long tails of the functions
$w_B^{Coul}(q)$ and $\tilde w_B^{'\;\,Coul}(q)$ are absent. It is
easy to see that the expressions (\ref{DC61b}),(\ref{DC62b}) in the
limit of small $q$ (the lower limit of the integrals in these
equations is taken equal to zero, which corresponds to the Landau
small-$q$ expansion) are appropriate in the Fokker-Planck equation
to the Landau approach for the kinetic equation for plasma. In
this case the coefficients $A_\alpha({\bf p})$ and
$B_{\alpha\beta}({\bf p})$ read
\begin{equation}
A_\alpha({\bf p})\cong \frac{p_\alpha}{3} \int d^3 q q^2 \tilde
w_B^{'\;\,Coul}(q)\cong -\frac{16 \pi^2 Z^2 e^4 p_\alpha}{3 M
T}\ln\left(\frac{q_{max}}{q_{min}}\right)J, \label{DC63b}
\end{equation}
\begin{equation}
B_{\alpha\beta}\cong \frac{\delta_{\alpha\beta}}{6} \int d^3 q q^2
w_B^{Coul}(q) \cong \delta_{\alpha \beta} \frac{16\pi^2 Z^2
e^4}{3} \ln\left(\frac{q_{max}}{q_{min}}\right) J, \label{DC64b}
\end{equation}
\begin{equation}
J=\int_0^\infty du\,u\, f_b (u^2)=\frac{n \sqrt \mu}{(2\pi)^{3/2}
\sqrt  T}. \label{DC65b}
\end{equation}
Therefore, one can rewrite $A_\alpha\equiv - p_\alpha \nu_{ie}$,
where $\nu_{ie}$ is the characteristic frequency friction ions on
electrons:
\begin{equation}
\nu_{ie}= \frac{4 \sqrt{2 \pi} Z^2 e^4 n \mu^{1/2}}{3 M T^{3/2}
}\ln\left(\frac{q_{max}}{q_{min}}\right).\label{DC66b}
\end{equation}
The corresponding friction force per unit volume $F_{ie}$ is equal 
${\bf F_{ie}}=n_i M {\bf U}\nu_{ie}$, where ${\bf F_{ie}}$
is the relative velocity of the electrons and ions [27].
In fact, the divergence at large $q$ handled by a cut-off are not an artefact.
This becomes clear when calculating the equilibrium
function $w_B^{Coul,\;0}({\bf q, p})$ more accurately, without expansion on
small values of $q$.

For the equilibrium distribution function $f_b^0(u)=n_e (\mu/2\pi
T)^{3/2}exp(-\mu u^2/2T)$ the PTF function reads
\begin{eqnarray}
w_B^{Coul,\;0}({\bf q, p}) =\frac{4 n_e Z^2
e^4\pi}{q^5}\exp[-\mu(v^2-{\bf q \cdot v} /\mu)/2T]
\int_{q/2\mu}^\infty du^2\,
 (\mu/2\pi T)^{3/2}\exp[-\mu u^2/2T]\nonumber\\
=\frac{4n_e Z^2 e^4 \mu^{1/2}}{\sqrt{2\pi T} q^5}
\exp[-\mu(v^2-{\bf q \cdot v} /\mu+q^2/4\mu^2)/2T]
 . \label{DC67b}
\end{eqnarray}
\begin{eqnarray}
w_B^{'\,Coul,\;0}({\bf q, p})=\frac{2 n_e Z^2 e^4
\mu^{1/2}}{\sqrt{2\pi M T^3} q^5} \exp[-\mu(v^2-{\bf q \cdot v}
/\mu+q^2/4\mu^2)/2T]
 . \label{DC68b}
\end{eqnarray}
Using the Fokker-Planck approximation for the coefficients
$A_\alpha({\bf p})$ and $B_{\alpha\beta}({\bf p})$, and
Eqs.~(\ref{DC67b}),\,(\ref{DC68b}) we find
\begin{equation}
A_\alpha({\bf p})\cong \frac{p_\alpha}{3} \int d^3 q q^2 \tilde
w_B^{'\,Coul,\;0}({\bf q, 0})= \frac{p_\alpha 2 n_e Z^2 e^4
\mu^{1/2}}{3 \sqrt{2\pi M T^3}} J_1\;,\label{DC63b}
\end{equation}
\begin{equation}
B_{\alpha\beta}\cong \frac{\delta_{\alpha\beta}}{6} \int d^3 q q^2
w_B^{Coul,\;0}({\bf q, 0})=\frac{2n_e Z^2 e^4
\mu^{1/2}\delta_{\alpha\beta}}{3 \sqrt{2\pi T}}J_1,  \label{DC64b}
\end{equation}
where
\begin{eqnarray}
J_1=\int d^3 q q^{-3} \exp[-q^2/8\mu T]=4 \pi \int^\infty_0
\frac{d q}{q} \exp[-q^2/8\mu T]\simeq 4 \pi \int^\infty_{q_{min}}
\frac{d q}{q} \exp[-q^2/8\mu T]\nonumber\\= 2 \pi
\int^\infty_{\frac{q^2_{min}}{8\mu T}} \frac{d\zeta}{\zeta}
\exp[-\zeta]=
 -2\pi Ei(-q^2/2 \mu T)|_{q_{min}}^\infty \simeq
 -2\pi Ei(-q_{min}^2/2\mu T). \label{DC65b}
\end{eqnarray}
We can suppose that the minimal momentum transfer  $q_{min}$ is
determined from the equality $q_{min}^2/2\mu T=r_{min}/r_{max}$.
According to the Landau theory for a weakly interacting plasma
$r_{min}/r_{max}=Ze^2/T r_D \ll 1$ for $Ze^2/\hbar v_T \gg 1$, or
$r_{min}/r_{max}=\hbar^2/2 \mu T r_D \ll 1$ for the opposite
inequality $Ze^2/\hbar v_T \ll 1$. Here $r_D$ is the Debye radius
and $v_T=\sqrt {T/\mu}$ is of the order of the thermal velocity. In
our approach the cut-off for the small momenta is satisfied
automatically and corresponds to the second inequality (the
"quantum" case).

For a weakly non-ideal plasma this means a cut-off at the minimal
momentum $q_{min}=\hbar/r_D$. Then
\begin{eqnarray}
J_1=-2 \pi C + 4 \pi \ln \left(\frac {r_D}{r_{min}}\right) \simeq
4 \pi \ln \left(\frac {r_D}{r_{min}}\right), \label{DC67b}
\end{eqnarray}
where $r^2_{min}\equiv \hbar^2/2 \mu T$ and $C \simeq 0.577$ is the
Euler constant.

It is easy to verify that $\tilde W'(q)= W(q)/2MT$ and $\tilde
W''(q)=W(q)/4M^2 T^2$ (in the case under consideration $W({\bf q,
p})\equiv w_B^{Coul,\;0}({\bf q, p})$). Therefore, for the
equilibrium case the usual Fokker-Planck equation for heavy
particles (ions or dusty particles in  dusty plasmas) is,
naturally, valid with a good accuracy, owing to the exponential
convergence of the integrals in the coefficients $A(s)$ and
$B_\alpha(s)$ is provided at high values of $q$. The term with
$W''$ in Eq.~(\ref{DC7b}) is negligible, according to the above
general statement. However, for small $q$ the coefficients $A$
and $B_\alpha$ have the logarithmical divergence typical for
Coulomb interaction because $W\sim 1/q^5$ just like $W'(q)$. As
follows from Eq.~(\ref{DC58b}), this divergence not
only exists for equilibrium, but for an arbitrary distribution function
$f_b$. The simplest physical way to avoid this divergence is to
cut the integrals for $A$ and $B_\alpha$ in Eqs.~(\ref{DC17b}),
(\ref{DC19b}) for small $q$ by the Debye radius $1/r_D$, following
the well known the Landau procedure. We are more interested to
find examples for non-exponential behavior of $W$, which may
occur, e.g., for some specific non-Maxwellian distributions
$f_b$.

\section{Models of anomalous diffusion for Coulomb interaction}

Now we can calculate the coefficients for models of anomalous
diffusion in plasma-like systems.

At first we calculate the model of a Coulomb system with two species of particles
with masses $\mu$ and $M\gg \mu$.
Let us suppose that in the model under consideration the small
particles are described by a prescribed stationary distribution
$f_b=n_b \phi_b /u_0^3$ (where $\phi_b$ is the non-dimensional
distribution, $u_0$ is the characteristic velocity for the
distribution of the small particles) and $\xi \equiv (u^2+
v^2-{\bf q\cdot v} /\mu)/u_0^2$.
\begin{eqnarray}
W_a ({\bf q, p})= \frac{8\pi Z^2 e^4 n_b}{u_0 q^5}
\int^\infty_{(q^2/4\mu^2+v^2-{\bf{q\cdot v}}/\mu)/u_0^2} d\xi
\,\cdot \phi _b (\xi).\label{DC30b}
\end{eqnarray}

First, let us consider a power type distribution $\phi_b(\xi)=C/\xi^\gamma$ ($\gamma>1$)
\begin{eqnarray}
W_a({\bf q, p})=  \frac{8\pi Z^2 e^4 n_b C}{u_0 q^5}\frac{\xi^{1-\gamma}}{(1-\gamma)}|_{\xi_0}^\infty=
\frac{8\pi Z^2 e^4 n_b C}{u_0 q^5}\frac{\xi_0^{1-\gamma}}{\gamma-1}
,\label{DC31b}
\end{eqnarray}
where $\xi_0\equiv (q^2/4\mu^2+v^2-{\bf{q\cdot v}}/\mu)/u_0^2$.

For the case $p=0$ the value $\xi_0 \rightarrow \tilde \xi_0
\equiv q^2/4\mu^2 u_0^2$ and we arrive at the following expression for
anomalous $W \equiv W_a$:
\begin{eqnarray}
W_a ({\bf q, p=0})=\frac{8\pi Z^2 e^4 n_b  C}{u_0 q^5}\frac{(q / 2\mu u_0)^{2-2 \gamma}}{\gamma-1}=\frac{2^{2\gamma+1}
\pi Z^2 e^4 n_b u_0^{2\gamma-3}\mu^{2\gamma-2}C}{(\gamma-1)q^{2\gamma+3}}
.\label{DC32b}
\end{eqnarray}

To determine the structure of the transport process and the
kinetic equation in velocity space we have to determine also the functions
$\tilde W'(q)$ and $\tilde W''(q)$.

If $p\neq 0$ we have to
use the full expression $\xi_0\equiv (q^2/4\mu^2+p^2/M^2-{\bf{q\cdot
p}}/M \mu)/u_0^2$ and it derivatives on ${\bf q \cdot p}$ at
$p=0$: $\xi'_0=-1/M \mu u_0^2$ and $\xi''_0=0$. Then
\begin{eqnarray}
\tilde W'({\bf q, p})\equiv \frac{8\pi Z^2 e^4 n_b C}{M \mu u^3_0 q^5 \xi_0^{\gamma}};
 \;\; \; \tilde W''({\bf q, p})\equiv \frac{8\pi \gamma Z^2 e^4 n_b C}{M^2 \mu^2 u^5_0 q^5 \xi_0^{\gamma+1}}. \label{DC34b}
\end{eqnarray}
For $p=0$ ($\xi_0 \rightarrow \tilde \xi_0$) we obtain
the functions
\begin{eqnarray}
\tilde W'(q)\equiv \frac{2^{2\gamma+3}\mu^{2\gamma-1}u_0^{2\gamma-3}\pi Z^2 e^4 n_b C}{M q^{2\gamma+5}};
 \;\; \; \tilde W''(q)\equiv \frac{2^{2\gamma+5}\gamma \mu^{2\gamma}u^{2\gamma-3}_0\pi Z^2 e^4 n_b C}{M^2  q^{2 \gamma+7}}.  \label{DC35b}
\end{eqnarray}

For the function $A(s)$, according to Eq.~(\ref{DC17b}), we find
\begin{eqnarray}
A(s)\equiv 4\pi \int_0^\infty dq q^2 \left[\frac{sin\, (q s)}{q
s}-1\right]W(q)= 4\pi C_a
 \int_0^\infty d q
\frac{1}{q^{2\gamma+1}} \left[\frac {sin(q s)}{qs}-1\right].
\label{DC33b}
\end{eqnarray}

Comparing the reduced equation (see below) in velocity space
with the diffusion in coordinate space
($2\gamma+3\leftrightarrow\alpha$ and $W(q)=C_a/q^{2\gamma+3}$) we
can conclude the integral in the right-hand
side of Eq.~(\ref{DC33b}) (3d case) converges if $3<2\gamma+3<5$
or $0<\gamma<1$. The inequality $\gamma<1$ implies
convergence for small $q$ ($q\rightarrow0$) and the inequality
$\gamma>0$ implies convergence for $q\rightarrow\infty$.
Likewise for the integral in $B(s)$:
\begin{eqnarray}
B(s)= \frac{4\pi}{s^2} \int_0^\infty dq q^2 \left[cos\, (q
s)-\frac{sin (q s)}{q s}\right]\tilde W'(q); \label{DC36b}
\end{eqnarray}
convergence of $B(s)$ exists for small $q$ if $\gamma<0$ and for
large $q\rightarrow\infty$ if $\gamma>-3/2$. Again, it is easy to
show that the term with $W''$ can be omitted.

Therefore, for convergence of $A$ and $B$ for a large $q$
we require convergence for $A$, which implies $\gamma>0$. For convergence for small $q$ it is sufficient to have
convergence for $B$, implying $\gamma<0$. Therefore, for the case of
pure power behavior of the function $f_b(\xi)$ convergence is
absent. It is also clear that the function $f_b(\xi)=C/\xi^\gamma$
($\gamma>1$) cannot be normalized. However, for
anomalous diffusion in momentum space in reality the
convergence for small $q$ is always obtained, e.g., by a finite
value of $v$ or by a change of the small $q$-behavior of $W(q)$ by
screening (compare with the examples of anomalous diffusion in
coordinate space [13]). Therefore, in the model under
consideration, the "anomalous diffusion in velocity space"  for
a power behavior $f_b(u)$ (and as a consequence with a power
dependence of $W(q)$ and $\tilde W'(q)$) on large $q$ exists if for
large $q$ the asymptotic behavior of $W(q\rightarrow\infty)\sim
1/q^{2\gamma+3}$ with $\gamma>0$. At the same time the expansion
of the exponential function in Eqs. (\ref{DC17b}),\,(\ref{DC19b})
under the integrals, leading to the Fokker-Planck type kinetic
equation, is invalid for power-type kernels $W(\bf {q, p)}$.

As an example of the above statements, let us consider the
Cauchy-Lorentz-like distribution for the function $f_b$
($r=3$):
\begin{eqnarray}
f_b (u^2) = n_b \frac{v_0}{\pi^2(u^2+v_0^2)^2}. \label{DC36ba}
\end{eqnarray}
Then we find
\begin{eqnarray}
W_a ({\bf q, p})=\frac{8 Z^2 e^4\pi}{q^5} \int_{q/2\mu}^\infty
du\,u\, f_b (u^2+ v^2-{\bf q \cdot v} /\mu)=\frac{4 v_0^3 Z^2
e^4}{\pi q^5} \nonumber\\\times\int_{\xi_0}^\infty d\xi\,
\frac{1}{1+\xi^2}= \frac{4 v_0^3 Z^2 e^4}{\pi
q^5}\left\{\frac{\pi}{2}-\arctan \xi_0 \right\},  \label{DC36bb}
\end{eqnarray}
where $\xi_0\equiv(q^2/4\mu^2+v^2-{\bf{q\cdot v}}/\mu)/v_0^2$ and
\begin{eqnarray}
\tilde W'_a ({\bf q, p})=\frac{4 v_0 Z^2 e^4}{M\mu\pi
q^5}\frac{1}{\xi_0^2+1}.\label{DC36bc}
\end{eqnarray}
For large $q$ the functions (\ref{DC36bb}) and (\ref{DC36bc}) tend to $W_a ({\bf q, p})\simeq 16
v_0^5\mu^2 Z^2 e^4/\pi q^7$ and $\tilde W'_a ({\bf q, p})\simeq 64
v^5_0 \mu^3 Z^2 e^4/M\pi q^9 $. For small $q$ convergence of the
coefficients $A(s)$ and $B(s)$ cannot be obtained since these
functions are determined by the expressions $W_a ({\bf q},0)$ and
$\tilde W'_a ({\bf q}, 0)$. However, this problem can be avoided
by using a cut-off of the respective integrals
(\ref{DC33b}) and (\ref{DC36b})  at small $q$ or by modification of the
distribution (\ref{DC36ba}) at small $q$ (in the spirit of the
respective cut-off for anomalous diffusion in coordinate space
[12]). For large $q$ the Cauchy-Lorentz-type of
distributions have long tails, thus leading to anomalous diffusion.

Let us now consider the formal general model for which we will
not connect the functions $W(q)$ and $\tilde W'(q)$ with a
concrete form of $W({\bf q, p})$. Therefore we consider the
problem suggesting some behavior of the function $W({\bf q, p})$, but not on
the level of the distribution function $f_b$. In general the functional
$W({\bf q, p})$ is unknown. In this case one can
suggest that, independently one from
another, the functions $W(q)$, $\tilde W'(q)$ and $\tilde W''(q)$ possess a power-type $q$-dependence for a large $q$.

As an example, this dependence can be taken as the power type for
two functions $W(q)\equiv a / q^\alpha$ and $\tilde W'(q)\equiv
b /q^\beta$, where $\alpha>0$ and $\beta$ are independent.
Then, as follows from the consideration above, convergence of
the function $W$ exists if \,$5>\alpha>3$ (for asymptotically
small and large $q$, respectively). For the function $\tilde W'(q)$
the convergence condition is $5>\beta>2$ for asymptotically small
and large $q$, respectively.

Finally for the function $\tilde W''(q)$ the convergence condition
is $7>\eta>5$ (for asymptotically small and large $q$,
respectively). In this case the terms with $W''$ can be omitted
(for the same reasons as above).

For this example the kinetic equation Eq.~(\ref{DC16b}) reads
\begin{eqnarray}
\frac{df({\bf s},t)}{dt} = P_0 s^{\alpha-3} f({\bf s},t)+
s^{\beta-5} P_1 s_i\frac{\partial}{\partial s_i} f({\bf s},t),
\label{DC38b}
\end{eqnarray}
where
\begin{eqnarray}
P_0= 4\pi a \int_0^\infty d\zeta \zeta^{2-\alpha}
\left[\frac{sin\, \zeta}{\zeta}-1\right], \label{DC39b}
\end{eqnarray}
\begin{eqnarray}
P_1= 4\pi b \int_0^\infty d\zeta \zeta^{2-\beta} \left[cos\,
\zeta-\frac{sin \zeta}{\zeta}\right]. \label{DC40b}
\end{eqnarray}

Taking into account the isotropy in $s$-space we can
rewrite Eq.~(\ref{DC38b}) in the form
\begin{eqnarray}
\frac{df(s,t)}{dt} = P_0 s^{\alpha-3} f(s,t)+ s^{\beta-4} P_1
\frac{\partial}{\partial s} f(s,t), \label{DC42b}
\end{eqnarray}

Naturally, Eqs.~(\ref{DC38b}) and (\ref{DC42b}) can be formally
rewritten in momentum (or in velocity) space via the fractional
derivatives of various orders (see below). Therefore, as is easy
to see, for the purely power behavior of the functions  $W(q)$ and
$\tilde W'(q)$ the solution with the convergent coefficients
exists for powers in the intervals mentioned above. We
establish that the universal type of anomalous diffusion in
velocity space in the case under consideration exists if
$5>\alpha>3$, $5>\beta>2$. This universality seems similar to the universality
of the Levy distribution in coordinate space, where the power $\alpha$ of the dependence
of the PT-function in coordinate space, $W(\tilde q)\sim C/\tilde q^\alpha$ on the displacement $\tilde q$
lies in the interval \, $0<\mu\equiv\alpha-r<2$ ($r$ is the dimension of the coordinate space).

As is easy to see the stationary solution of  Eqs.~(\ref{DC42b}) reads
\begin{eqnarray}
f(s) = C' \exp\;\left[-\frac {P_0 s^{\alpha-\beta+2}}{(\alpha-\beta+2) P_1}\right]. \label{DF42baa}
\end{eqnarray}

Of course, the general description above is also valid for the more
complicated functions $W$ and $\tilde W'$, possessing a
non-power dependence on $q$ at small $q$ and an asymptotical power dependence on $q$ at large $q$.
In this case the limitations for convergence are connected only with large
values of $q$, namely it is enough to provide the inequalities $\alpha>3$ and $\beta>2$.
Simple examples of such type of PT-functions are (in analogy with anomalous diffusion in coordinate space [13]):
\begin{eqnarray}
W_a ({\bf q})=\frac{1-\exp(-\gamma q^n)}{q^\alpha}, \,\;\;  \tilde W'_a ({\bf q})=\frac{1-\exp(-\delta q^m)}{q^\beta} .\label{DC42ba}
\end{eqnarray}

The correspondig kinetic equations in these cases cannot be written in partial derivatives and
evolution of the system has to be described by Eq. (\ref{DC16b}), or for the isotropic case by (\ref{DC22b}).
If external forces are present, they have be included in the usual way in the left side of Eq. (\ref{DC16b}).
Physically, this type of PT-function behavior can appear, in
particular, for the case of a turbulent plasma, when the
development of some instability can create a strong chaotic
electrical field or irregular chaotic motion of one sort of particles with
a prescribed non-Maxvellian distribution function. In such a turbulent plasma
scattering with large transferring momenta can play a crucial
role.

\section{Representation in momentum space and connection with the fractional differentiation approach}

As mentioned, in general Eq.~(\ref{DC16b}) cannot be written
in terms of fractional differentiation. It confirms that the approach to
anomalous diffusion based on Fourier-transformation of the PT-
functions, in the form applied in this paper (see also [13],[19])
is a more general way for the problems under consideration.

However, for the purely power-type dependence of the functions
$W(q)\equiv a / q^\alpha$ and $\tilde W'(q)\equiv b /q^\beta$,
where $\alpha$, $\beta$ are independent and satisfy the inequalities $5>\alpha>3$ and $5>\beta>2$,
Eq.~(\ref{DC38b}) is appropriate and can be represented after inverse
Fourier-transformation in the following form (with the fractional derivatives)
\begin{eqnarray}
\frac{df({\bf p},t)}{dt} =P_0 \Delta^{\nu} f({\bf p},t)+ P_1 \Delta^{\lambda}
\left(\lambda + \frac{\partial}{\partial \,{\bf p}}{\bf p}\right) f({\bf p},t), \label{DF1}
\end{eqnarray}
where $\nu \equiv(\alpha-3)$ ($2>\nu>0$); $\lambda=\beta-5$ ($2>\lambda>0$). Here we introduced the
fractional differentiation operator in the momentum space $\Delta^\lambda f({\bf p},t)\equiv \int
d{\bf s} s^{2\lambda} exp(-i{\bf ps}) f({\bf s},t)$.

Let us consider now formally a specific particular model of
anomalous diffusion, for which we assume a structure of the PTF
$W({\bf q, p})$ with a rapid (say, exponential) decrease
of the function $\tilde W'(q)$. Therefore, the exponential function
under the integrals in the coefficients $B(s)$ can be expanded,
implying $B(s)=B_0$ (or $\beta=5$ and $B_0\equiv P_1$ in the notations of Eq.~(\ref{DC38b})). At the same time the function $W(q)\equiv
a/q^\alpha$ has a purely power dependence on $q$.
The kinetic equation Eq.~(\ref{DC42b}) then reads
\begin{eqnarray}
\frac{df({\bf s},t)}{dt} = P_0 s^{\alpha-3} f({\bf s},t)+ B_0
s_i\frac{\partial}{\partial s_i} f({\bf s},t), \label{DF2}
\end{eqnarray}
or in momentum space, according to (\ref{DF1})
\begin{eqnarray}
\frac{df({\bf p},t)}{dt} = P_0 \Delta^{\nu} f({\bf p},t)- B_0
\frac{\partial}{\partial p_i} [p_i f({\bf p},t)]. \label{DF3}
\end{eqnarray}

Eq.~(\ref{DF3}) is similar to the corresponding equation in [22], where a model of the Langevin equation with
a constant friction frequency $\nu_0\equiv-B_0$ has been considered. In [23] a similar model with
$\nu=3/2$ has been applied to
estimate the fusion rate in a hot rarified plasma.

The stationary solution of Eq.~(\ref{DF2}) is:
\begin{eqnarray}
f(s) = C' exp\;[-\frac {P_0 s^{\nu}}{\nu B_0}]. \label{DF4}
\end{eqnarray}
The corresponding distribution in $p$-space is proportional to the Levy-type distribution $W(y, \alpha') \equiv (2 y/ \pi) \int_0^\infty dt\, t\, sin (\,y\,t) \exp(-t^{\alpha'})$:
\begin{eqnarray}
f(p) = C' \int d^3 s exp(-i{\bf ps})exp\;[-\frac {P_0
s^{\nu}}{\nu B_0}]\equiv \frac{4\pi C'}{p} \int^\infty_0  ds s sin
(ps) exp\;[-\frac {P_0 s^{\nu}}{\nu B_0}],\label{DF5}
\end{eqnarray}
with $y\,\equiv p\,(\nu B_0/P_0)^{1/\nu}$ and $\alpha' \equiv \nu$.
As an example for the case $\nu=1$ we find $f({\bf p})$
\begin{eqnarray}
f(p) = \frac{8\pi C'P_0}{B_0 [(p^2+4P_0^2/B_0^2)]^2}.\label{DF5a}
\end{eqnarray}
In the case $\nu=1$ the long tail of the distribution is proportional to ~$p^{-4}$ and the distribution $f(p)$ corresponds with the Cauchy-Lorentz distribution. Normalization of the distribution $f(p)$ leads to the value $C'=n/(2\pi)^3$, where $n=N/V$ is the average density of particles undergoing diffusion in velocity space. A similar appoaach can be taken for other types of anomalous
diffusion in velocity space.

\section{Conclusions}
In the present paper the problem of anomalous diffusion for plasma-like
systems in momentum (velocity) space is consequently analyzed.
A new kinetic equation for anomalous diffusion in velocity
space has been derived recently in [19], without suggesting any stationary
equilibrium distribution function. We applied this
equation to a system of charged particles with different masses
to describe diffusion of heavy particles (ions, charged
grains) in the surrounding light particles (electrons for the
electron-ion plasma, electrons and ions for dusty plasmas). The
distribution of the light particles can be non-Maxwellian, which is
the cause for appearance of the long tails in the probability
transition function. Conditions of convergence for
the coefficients of the kinetic equation have been derived for a number of 
particular cases. It is found that a wide variety of anomalous processes in
velocity space exists.

In general the Einstein relation for
such a situation is not applicable, because the stationary state may
be far from equilibrium. For the case of normal diffusion the friction and
diffusion coefficients have been found explicitly for the
non-equilibrium case. For the equilibrium case the usual Fokker-Planck
equation in plasma is reproduced as a particular case.

\section*{Acknowledgment}
The authors are thankful to A.M. Ignatov and A.G. Zagorodny for the valuable discussions of
some problems, reflected in this paper. S.A.T. would like to thank
the Netherlands Organization for Scientific
Research (NWO) for support of his investigations on the problems of
stochastic transport in gases, liquids and plasmas.

\end{document}